\documentclass{article}
\usepackage[margin=1.5in]{geometry}
\usepackage{indentfirst}
\usepackage{amsmath}
\usepackage{float}
\usepackage{wasysym}
\usepackage{graphicx}
\usepackage{cleveref}

\title{Argon bubble formation in tantalum oxide-based films for gravitational wave interferometer mirrors}
\author{Rebecca B. Cummings$^1$, Riccardo Bassiri$^2$, Iain W. Martin$^1$ and Ian MacLaren$^1$}
\date{$^1$Department of Physics and Astronomy, The University of Glasgow, Glasgow, G12 8QQ, UK\\%
		$^2$E. L. Ginzton Laboratory, Stanford University, Stanford, California 94305, USA\\}

\begin{document}
	
	\maketitle
	
	\section*{Abstract}

	The argon content of titanium dioxide doped tantalum pentoxide thin films was quantified in a spatially resolved way using HAADF images and DualEELS. Films annealed at 300\;$^{\circ}$C, 400\;$^{\circ}$C and 600\;$^{\circ}$C were investigated to see if there was a relationship between annealing temperature and bubble formation. It was shown using HAADF imaging that argon is present in most of these films and that bubbles of argon start to form after annealing at 400\;$^{\circ}$C and coarsen after annealing at 600\;$^{\circ}$C. A semi-empirical standard was created for the quantification using argon data from the EELS atlas and experimental data scaled using a Hartree Slater cross section. The density and pressure of argon within the bubbles was calculated for 35 bubbles in the 600\;$^{\circ}$C sample. The bubbles had a mean diameter, density and pressure of 22\;\AA, 870\;kg/m$^3$ and 400\;MPa, respectively. The pressure was calculated using the Van der Waals equation. The bubbles may affect the properties of the films, which are used as optical coatings for mirrors in gravitational wave detectors. This spatially resolved quantification technique can be readily applied to other small noble gas bubbles in a range of materials. 
	
	\section{Introduction}
	
	In recent years, gravitational waves from merging binary black holes \cite{Abbott2016, Abbott2016_2, Abbott2017_2, Abbott2017_3,Abbott2018} and binary neutron stars \cite{Abbott2017} have been detected by the Advanced LIGO \cite{Abbott2016_3} and Advanced Virgo~\cite{Acernese2015} gravitational wave detectors. These detectors are modified Michelson interferometers that measure changes, induced by gravitational waves, in the relative separation of mirrors located at the end of km-scale perpendicular arms. The interferometer mirrors have highly-reflective multi-layer optical coatings. Thermal noise arising from thermally-induced vibrations in the coatings is a critical sensitivity limit to current detectors at their most sensitive frequencies \cite{Abbott2016_3}. Coating thermal noise is related to the mechanical loss, or internal friction, of the coating materials. Reducing the coating loss is essential to enable the construction of more sensitive detectors, which would be able to detect more gravitational waves from a wider range of astrophysical sources. The current high index layer in the LIGO and Virgo mirrors is amorphous Ta$_2$O$_5$, doped with TiO$_2$ \cite{Abbott2016,Acernese2015,Granata2016}, deposited using argon ion beam assisted deposition \cite{Harry2006}. Fully understanding the atomic and nano structure of the mirror coatings is the first step in reducing their mechanical loss. 

	Recent studies have shown that there are voids in the tantala films \cite{Anghinolfi2013} and also, based on fitting of ellipsometry data and Rutherford backscattering (RBS) measurements, that the films contain argon \cite{Prasai2019}. Subsequent work on hafnia thin films showed in those samples that argon lies within the bubbles \cite{Harthcock2019} and in tantala films it was shown that bubbles form on annealing \cite{Fazio2020}. Little was known, however, on the form of the argon inside the bubbles, or to what extent they are affected by annealing. 
	
	 Bubbles of noble gas in materials have been well documented over the years, mainly from ion implantation into solid materials. They are of particular interest for materials used in nuclear reactors, where fission products such as alpha particles and heavier nuclei will implant into the walls of the reactor. Once there, the alpha particles capture electrons and become helium nuclei. The heavier nuclei, especially xenon and krypton, will do the same. Noble gases do not interact strongly in a chemical fashion with their surroundings, so their main effect is local strain. If there are multiple noble gas atoms within a diffusion path length of each other the strain field interaction will cause them to cluster together. If there are enough of them they can form nanoscale or larger bubbles in the material \cite{Barnes1960, Barnes1963, Greenwood1959}. 
	 
	 Studying nanoscale gas bubbles in materials requires an imaging method with sufficient resolution to determine the number of interior gas atoms. Transmission electron microscopy (TEM) is such one method but scanning transmission electron microscopy (STEM) offers particular advantages. If the gas inside the bubble has a significantly different density to the material surrounding it, then HAADF imaging (only available in STEM) will show Z-contrast \cite{Pennycook1991}. This contrast is an easy way of telling where there are different elements within the sample. STEM also offers analytical techniques like EDX \cite{Nogita1998} and EELS \cite{Klimiankou2005} and there has been significant work in recent years on absolute quantification using both EDX \cite{Watanabe2006, MacArthur2016} and EELS \cite{Craven2016,Craven2018}. These techniques have been used to study the density, structure and chemistry of the noble gas bubbles \cite{Blackmur2018, Nogita1998,Izui1984}. There are fewer studies on argon bubbles using TEM as argon is less common as a fission product in nuclear reactorsbut some do exist \cite{Vasiliu1975, Klimiankou2005}.
	
	The current paper uses a combination of HAADF imaging and EELS to study argon in the Ta$_2$O$_5$-based coatings; quantifying the argon content in a spatially resolved way. This quantification allows calculations of the density and pressure of the argon within the bubbles. 
		
	Previous work done at Glasgow showed that absolute quantification using EELS is possible using a technique called DualEELS \cite{Gubbens2010, Scott2008} and an experimental standard for the material being quantified. Quantifying gaseous argon meant the latter presented a serious challenge. There is, however, published EELS data for argon gas \cite{eelsatlas}, which will give the correct edge shape for the pressure used in that experiment. Additionally, it is well known that the Hartree-Slater calculations of cross sections for L$_{2,3}$ edges of lighter elements are reasonably accurate, so long as the measurement is made at a high enough energy above the edge onset \cite{Egerton1993, Scott2008}. The solution, therefore, to the lack of an experimental standard was found by splicing together the absolute cross section from the Hartree-Slater calculation at a suitably high energy above the edge and the edge shape from the EELS atlas data \cite{eelsatlas}. This produced a semi-empirical cross section that could be used to fit real spectrum images (SI) of materials containing argon. Moreover, since it is an absolute cross section, the fitting coefficients are quantitatively meaningful \cite{Craven2018}, which means that the number of argon atoms in every pixel of an SI can be counted and allows for the quantitative mapping of argon at sub-nanometre spatial resolution.
	
	\section{Methods}
	\subsection{Sample preparation}
	
	The samples in this study were part of a series produced at CSIRO using argon ion beam assisted deposition onto fused silica substrates. They are nominally 75\% Ta$_{2}$O$_{5}$ and 25\% TiO$_2$ in stoichiometry, although characterisation using a similar technique to this paper found they were closer to only 10\% TiO$_2$ \cite{Isa2019}. Some of the samples were heat treated in air at a range of temperatures (300, 400 and 600\;$^{\circ}$C) for 24h. These were prepared for transmission electron microscopy observation using a conventional cross section preparation method as described by Hart \cite{Hart2016}.
	
	\subsection{Data collection}
	
	STEM observation was performed using a JEOL ARM200F operated at 200\;kV, and equipped with a Gatan GIF Quantum ER spectrometer. HAADF images were used to select areas of interest and to analyse the physical structure of the sample. EELS spectrum imaging was used to collect data on the composition at high spatial resolution. In all cases, a convergence angle of 29\;mrad was used for the probe. In the EELS, a camera length was chosen to give a collection angle into the spectrometer of 36\;mrad. Dual range EELS was used, to collect both the low loss and the high loss together using times of 0.000497238\;s and 0.0895028\;s and energy offsets of 0\;eV and 180\;eV, and all data was recorded at a dispersion of 0.5\;eV/ch.
	
	\subsection{Data processing}
	
	The EELS datasets were processed using Gatan Digital Micrograph (DM) using similar methods to those described in previous publications \cite{Craven2016, Bobynko2015}. The low- and high-loss SI were first aligned by the zero-loss peak of the low-loss image and then any x-rays that caused a spike larger than 10$\sigma$ above background were removed. The energy range of the low-loss datasets was cropped to 150\;eV so only channels containing useful information remained. The SIs were then run through a principal component analysis plug-in \cite{Lucas2013} in order to separate real signal from random noise. The high and low-loss spectra were spliced together and Fourier-logarithm deconvolution \cite{Egerton_eels} was performed to yield a single scattering distribution. This is essential to in order to produce model-based fitting with consistent shapes, even if there are thickness variations in the area analysed \cite{Craven2018}.
	
	A semi-empirical cross-section of the Ar-L$_{2,3}$ edge was constructed from the EELS atlas \cite{eelsatlas} edge shape, which has a dispersion of 0.3\;eV/channel. First, this dataset was re-binned to 0.5\;eV/ch to match the experimental data. This was done using a script based on ideas developed in separate work, which is on the correction of dispersion non-linearities in some EELS data \cite{Webster2020}. The Hartree-Slater cross section for the Ar-L$_{2,3}$ edge was calculated for the experimental conditions used in our analysis. This cross section was then used to scale the experimentally measured argon edge to give a tail of the correct magnitude, to allow for its use in fitting to the experimental data (\cref{spectra_bub})a)).
	
	After this, the experimental datasets were fitted using the multiple linear least squares (MLLS) method as implemented by the standard function in DM. This function fits an experimental SI as the sum of two or more standard spectra and requires several datasets as references. The argon edge cross section was calculated as described above. The tantala-titania glass standard was estimated by finding an area in the spectrum image where there was little or no argon. This process is straightforward as the Ar-L$_{2,3}$ edge has a strong peak at 250\;eV, which is above the edge onset for the Ta-N$_{4,5}$ edge (226\;eV) \cite{Harry2006, Isa2019}, but below the Ti-L$_{2,3}$ (454\;eV). In some cases, two such spectra were recorded if the area scanned contained a significant thickness variation in the sample, in order to remove any thickness effects. Ultimately, there was no effort to determine Ti-Ta stoichiometry, which has been done previously \cite{Harry2006, Isa2019}. This means that the only standard required to have a quantitative meaning is the argon one.
	
	The MLLS fit (\cref{mlls}) was restricted to 220-280\;eV so as to not include the carbon K-edge at 284\;eV, since C contamination on surfaces could vary with position and skew results. The fitting was an iterative process as it was not always clear from the original image where the boundaries of the bubbles were, and areas chosen for selection of components were adjusted until the fit residuals were minimised. This iterative approach is based one the used by Annand \cite{Annand2015}.
	
	As stated above, the fit coefficients are quantitatively related to the number of atoms illuminated by the beam in each pixel of the SI and so can be used to calculate the areal density of the argon atoms in the bubbles:
	
	\begin{equation}
		\label{dI}
		{\frac{dI}{dE}} = I_0 \sum_{\text{(elements\;i)}}\sum_{\text{(shells\;j)}} N_{j} \frac{d\sigma_{ij}}{dE},
	\end{equation}
	
	where $\frac{dI}{dE}$ is the intensity per unit energy at a given energy loss, and $\frac{d\sigma_{ij}}{dE}$ is the partial cross section of an electron in the $j$th shell of the $i$th atom; as detailed and used in \cite{Craven2018}. Specifically, in this case
	
	\begin{equation}
		\label{n_ar}
		N_{Ar} = \frac{I_{Ar}}{I_{0} \sigma (\Delta E)},
	\end{equation}
	
	where $I_{Ar}$ is the spectrum intensity of the argon edge in the energy range, $\Delta E$, used in the fit, $I_0$ is the intensity of the zero-loss peak in the low-loss data, and $N_{Ar}$ is the number of argon atoms per unit area. The number of atoms in each column represented by a pixel in an image was calculated by multiplying $N_{Ar}$ by the area of the pixel.
	
	As in previous work, the differential cross section, $\frac{d\sigma_{ij}}{dE}$ is in units of Barns/eV \cite{Craven2018}. In order to convert it to a total cross section, $\sigma(\Delta E)$, this needs to be integrated over the energy range $\Delta E$ and multiplied by the channel width. 
	
	The bubbles were assumed to be spherical, based on their appearance in the HAADF images. The diameter of each bubble was taken to be the horizontal diameter from the fit coefficient maps. This horizontal value was taken rather than the vertical because of the drift that can occur when the beam rasters across the sample. It has little effect along the line but between the top and the bottom of the bubble it has a much greater effect, potentially over 1\;nm. 
	
	A border of one pixel either side of the bubble was also included, based on comparison between elemental maps of argon and HAADF images of the same object. This is because the argon content in the outer part of the bubble will appear negligible because of the very short beam path through the bubble. Not including the border, however, will make the bubble too small and give unrealistically high densities of argon, consequently affecting the pressure calculations. The volume was calculated using the standard equation relating diameter and volume in a sphere.
	
	The pressure inside the bubbles was calculated using the Van der Waals equation of state \cite{VdW1873}:
	
	\begin{equation}
		\label{vdw}
		(P + \frac{an^{2}}{V^2})(V- nb)= nRT,
	\end{equation}
	
	where $P$ is pressure, $V$ is volume of gas, $n$ is the number of moles, $T$ is temperature, $R$ is the gas constant and $a$ and $b$ are experimentally derived constants unique to each material (1.345\;L$^2$ bar mol\;$^{-2}$ and 0.03219 L mol\;$^{-1}$, respectively, for argon \cite{cphandbook}). The ideal gas law could not be used as its assumptions do not hold in this case; at such small volumes and high densities the argon atoms cannot be assumed to be non-interacting point particles.
	
	\section{Results and Discussion}
	
	\begin{figure}[h!]
		\centering
		\includegraphics[width=0.9\linewidth]{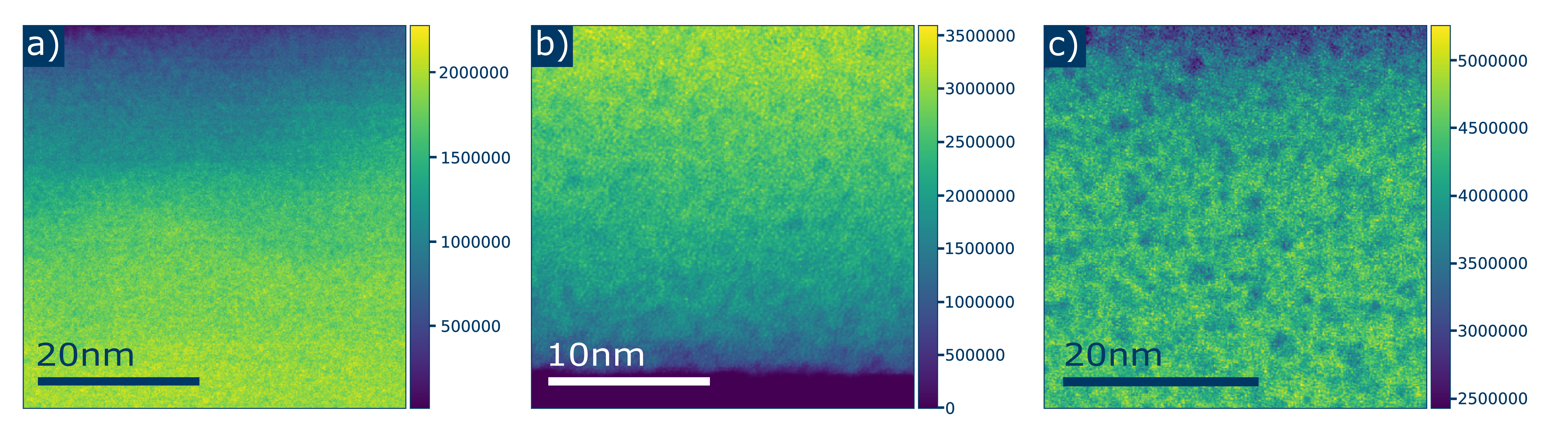}
		\caption{HAADF images of larger areas containing bubbles: a) the as-deposited sample showing no bubbles; b) the sample annealed at 400\;$^{\circ}$C for 24 hours, showing small bubbles, the strong contrast variation is due to this image being taken at the very edge of the sample; c) the sample annealed at 600\;$^{\circ}$C for 24 hours showing larger bubbles, with only a slight thickness increase from top to bottom. In both samples, most bubbles appear rounded, and it is clear there is some range in the bubble sizes in both cases. The colour map is \textit{viridis} from \textit{matplotlib} \cite{Matplotlib}, with the argon bubbles showing up as darker areas. The scale bars show the relative intensities of each image.}
		\label{haadf}
	\end{figure}
	
\begin{figure}[h!]
		\centering
		\includegraphics[width=0.9\linewidth]{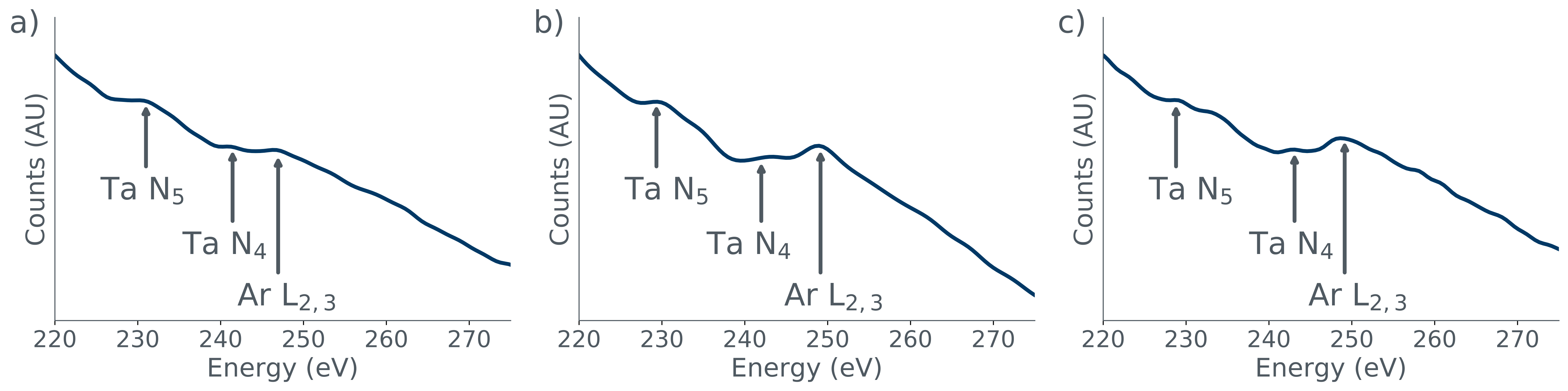}
		\caption{EELS spectra from three samples all annealed at different temperatures, showing clear evidence that argon is incorporated in all three. The relative intensities of the element edges are proportional to the amount of the element in the material. The tantalum N$_{5}$ (226\;eV onset) and N$_4$ (238\;eV onset) edges and the argon L$_{2,3}$ edge (248\;eV onset) are visible in all of the samples; a) the as-deposited sample; b) 400\;$^{\circ}$C annealed sample; c) the 600\;$^{\circ}$C annealed sample.}
		\label{ar edge strength}
	\end{figure}

	\Cref{haadf} shows HAADF images of thin areas of two samples, heat treated at 400\;$^{\circ}$C and 600\;$^{\circ}$C, respectively. In both cases lower-density (i.e. darker) regions are seen, all a few nm in diameter. They are mostly rounded in appearance, suggesting that they are approximately spherical voids or bubbles. They are clearly larger and more prominent in the sample annealed at the highest temperature. Similar features were seen in the sample heat-treated at 300\;$^{\circ}$C, but the images were very poor due to sample charging. They were smaller and fainter than those in the 400\;$^{\circ}$C sample. It would have been possible to recoat the 300\;$^{\circ}$C annealed sample with carbon to prevent the charging but as the bubbles in the 400\;$^{\circ}$C annealed sample were already difficult to analyse it was not deemed necessary. It was not possible to unambiguously resolve any such features in the non-annealed sample. It can be seen in \cref{ar edge strength}, however, that argon was still definitely present in these samples. The as deposited sample has a weak edge visible around 250\;eV and the edge for the 400\;$^{\circ}$C annealed sample is as strong as the one in the 600\;$^{\circ}$C annealed sample.
	
	While the exact sizes of the bubbles are hard to define from the unprocessed datasets it does appear that the average size of the bubbles grows with annealing temperature. Due to the poor signal-to-noise ratio in the 400\;$^{\circ}$C annealed datasets the full processing procedure was not possible. This means that the size distribution of the bubble diameter comes from the deconvolved-sliced images of the 600\;$^{\circ}$C annealed datasets only; as there were not enough clearly defined bubbles in the 400\;$^{\circ}$C annealed datasets to give a reliable distribution of bubble diameter. All data mentioned past this point stem from the 600\;$^{\circ}$C annealed sample.
	
	\begin{figure}[h!]
		\centering
		\includegraphics[width=0.8\linewidth]{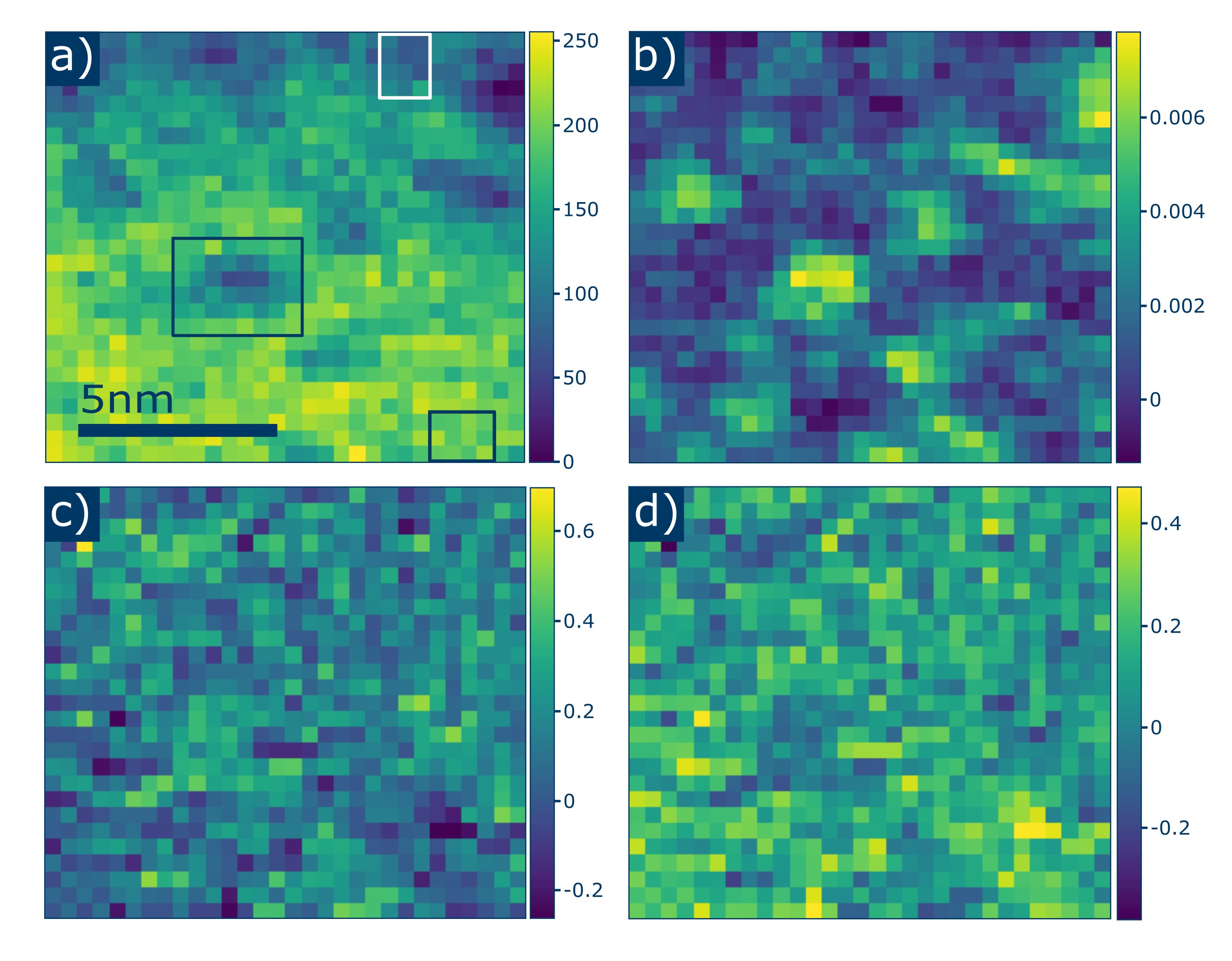}
		\caption{MLLS fit component maps for one of the datasets annealed at 600\;$^{\circ}$C; a) the fit to the single scattering distribution, with annotations to show where the matrix component spectra were taken from; top = thick matrix, bottom = thin matrix. The middle rectangle denotes an example of how the edges of the bubbles were drawn, with a border round the area of brightest intensity; b) the argon coefficient map, showing areas of intensities where the bubbles lie; c) thick matrix coefficient; d) thin matrix coefficient.}
		\label{mlls}
	\end{figure}

	The dataset of the MLLS fit shown in \cref{mlls} had a thickness gradient in the matrix, which is why there are two matrix components; a thick coefficient and a thin one. It seems the matrix was thicker at the top of the area scanned (as shown in \cref{mlls}). Not all samples displayed this gradient. 
	
	\begin{figure}[h!]
		\centering
		\includegraphics[width=0.9\linewidth]{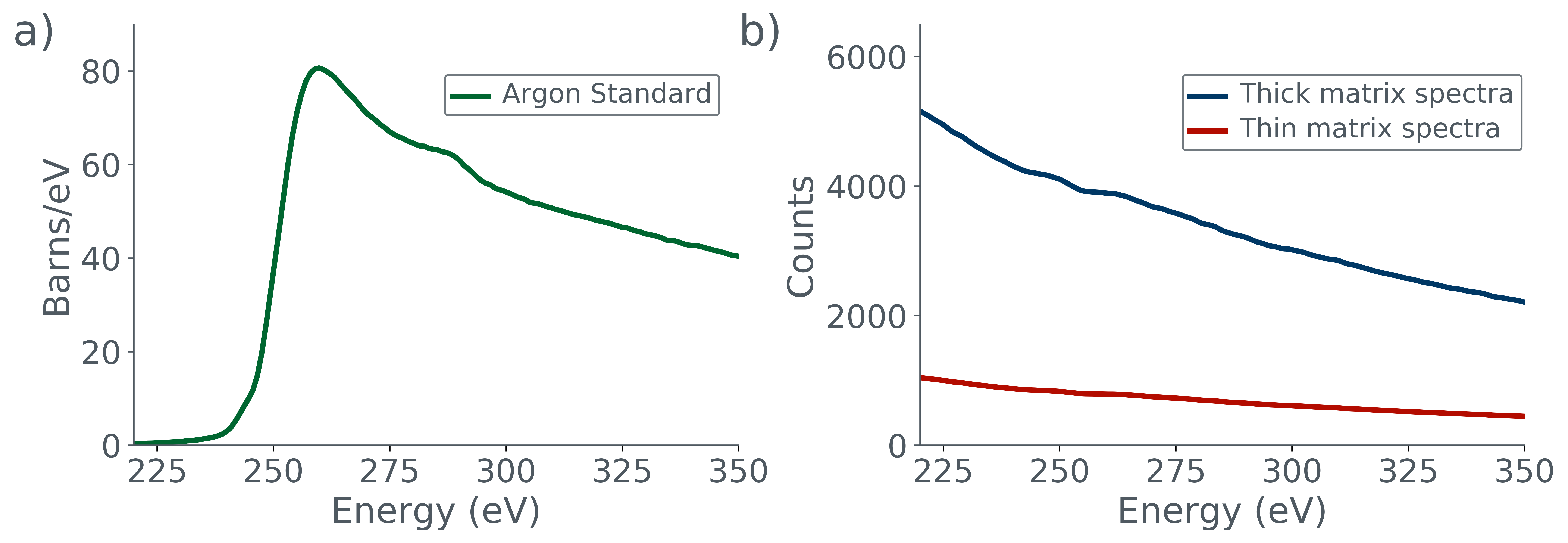}
		\caption{The contributions of the components of the MLLS fit for one bubble (i.e. the standard spectrum multiplied by the fit coefficient); a) for argon, based on the background-subtracted cross section; b) for thin and thick matrix components. Note the difference in the magnitude of the y-axes scale, the matrix components dominate, even in a bubble area, and the signal from the argon in the bubble is always a tiny fraction of the total EELS signal.}
		\label{spectra_bub}
	\end{figure}

	As stated, the fitting was an iterative process. \Cref{spectra_bub} shows the contributions to the overall spectrum from a bubble inside each of the three different components: the background-subtracted argon cross section (a), and the thin and thick matrix standard spectra (b). These spectra make it clear that even with an obvious bubble, the argon signal itself is a tiny fraction of the total EELS signal in this energy range. It is only through the intrinsic noise-reducing properties of the MLLS fit (which fits through the random noise) that such weak features could be reliably extracted and quantified. These spectra were calculated by multiplying the fit coefficient for the area by the standard spectrum. 
	
	\Cref{mlls} shows the actual maps of the fit coefficients for the three components shown in \cref{spectra_bub}.  The fit coefficient for the argon is of particular interest, as this can be directly interpreted into an absolute number of argon atoms per column (\cref{dI,n_ar}). From this the argon density inside of the bubbles can be calculated, as well as the pressure (\cref{vdw}).
	
	The matrix coefficients were made using spectra extracted from the single scattering distribution rather than a standard, so it is not possible to determine the absolute number of atoms from them. 
	
	\begin{figure}[h!]
		\centering
		\includegraphics[width=\linewidth]{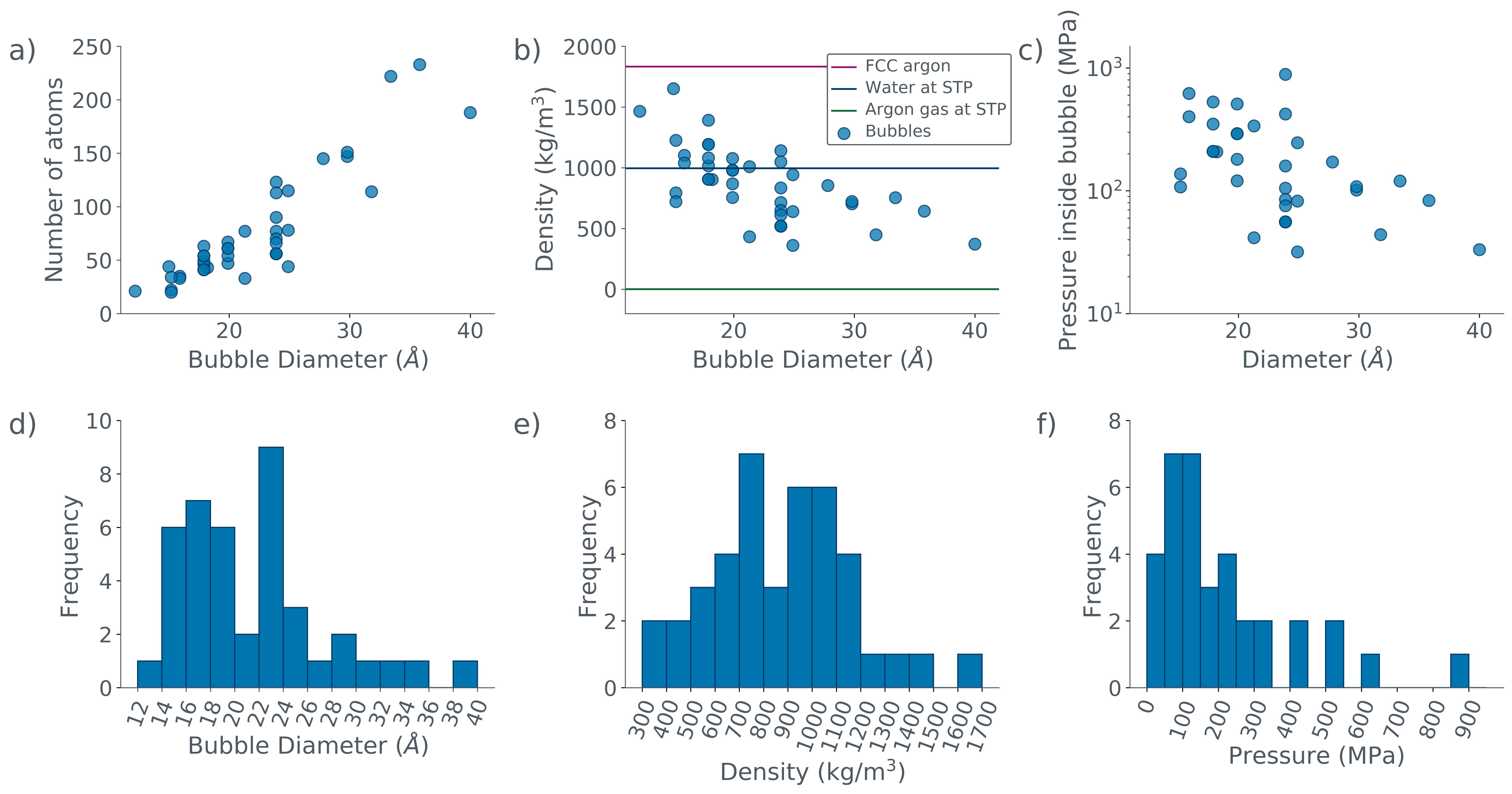}
		\caption{Analysis of the argon bubbles in the 600\;$^{\circ}$C sample; a) number of argon atoms in bubble against their diameter; b) density of the argon bubbles, with lines representing the density of argon gas at standard pressure and temperature (STP), water and argon arranged in the tightest possible face-centered cubic (FCC) packing; c) pressures of the bubbles against their diameter; d) distribution of bubble diameters; e) distribution of bubble densities; f) distribution of bubble pressures.}
		\label{ndp_graph}
	\end{figure}

	The number of atoms per bubble increases with bubble diameter, as might be expected. The precision with which the diameter can be measured is limited by pixel size (3-6\;\AA) and it is one potential cause of the wide range in the number of argon atoms in bubbles of the same size. It is hard to tell from the limited number of samples what the precise nature of the relationship between number of atoms and diameter is (linear, cubic etc). The bubbles that were especially numerically dense are most likely two bubbles laid over each other.  The diameters measured are in good agreement with those taken from HAADF images in previous studies \cite{Fazio2020}.
	
	For all of the bubbles that made up the final pressure analysis, the number of atoms per bubble is below the number that could fit assuming the argon was in crystal form (i.e. the maximum number of argon atoms that could physically fit in the bubble, ignoring temperature). FCC argon has a lattice parameter of 5.45\;\AA{ }at 81\;K, which translates to an atom spacing of 3.85\;\AA. The data suggests most of the bubbles are still in gas form but some of the smallest may be a high density fluid. Previous work on xenon and krypton suggests it can be solid at room temperature in bubbles formed from ion implantation \cite{Potter1989, Evans1985, Felde1984}. There is some evidence that argon can also form solid bubbles, with a lattice parameter around 5.15\;\AA \cite{Felde1984, Donnelly1986,Roussouw1985}.
	
	The bubbles are far less dense than the $\text{Ta}_{2}\text{O}_{5}$ surrounding them, which has a density of $\sim$7750kg/m$^3$ \cite{Isa2019}. This explains the HAADF contrast of \cref{haadf}, where the bubbles are darker than the surrounding $\text{Ta}_{2}\text{O}_{5}$ matrix.
	
	Using RBS, a similar sample was found to contain 3\% argon \cite{Prasai2019}. Comparing the number of atoms in the bubbles with the total number of atoms in the matrix (calculated using the number densities of atoms and mean free paths for inelastic scattering in $\text{Ta}_{2}\text{O}_{5}$ and $\text{TiO}_{2}$ \cite{Isa2019,Iakoubovskii2008}), results in a much lower argon contribution (\textless1\%). The remaining argon is likely still in the matrix as individual interstitial atoms or sub-nanometre clusters and not the bubbles, although it might continue to precipitate out into bubbles if a longer heat treatment were applied. Argon near the surface of these samples may also have been lost due to damage during sample preparation. A bubble volume fraction calculation was not done as it would require an argon-free $\text{Ta}_{2}\text{O}_{5}$ film with which to compare the argon-containing samples. Preliminary calculations suggest, however, that it may be around 4-5\%, on account of the relatively low density of the bubbles, when compared with the $\text{Ta}_{2}\text{O}_{5}$.
	
	As the density of the argon inside the bubbles is well above the density of argon at STP \cite{cphandbook}, the ideal gas law will not provide an accurate description of the pressure inside of the bubbles. The Van der Waals equation of state should describe the situation better at higher densities, and this put the average pressure inside of the bubbles on the order of 100\;MPa, well above the critical pressure of argon, which is approximately 5\;MPa \cite{McCain1967}. 
	
	This approach failed for the smallest bubbles (\textless20\;{\AA} diameter), giving either negative or unphysically high pressures. The Van der Waals equation is known to break down in extreme cases and if the smallest gas bubbles are high-density fluids, then a gas equation of state would not be an appropriate way to describe them. Nevertheless, the more likely explanation in these cases is that two bubbles were overlapped in the beam path through the specimen, apparently giving a number of atoms per bubble about double the expected level, which would automatically give unphysical density and pressure results. Recent work on xenon and krypton bubbles formed by ion implantation showed that the Van der Waals equation did not give an wholly accurate description of the pressures within the bubbles, and tended to overestimate them \cite{Jelea2020}.
	
 	The diameter and number of atoms were counted for 41 bubbles in total, the pressures calculated for 35 (since 6 had to be excluded for unphysically high densities, as accounted for above). The pressure inside the bubbles seems to decrease with bubble diameter but the large standard deviation makes it difficult to assess the exact nature of the relationship between diameter and pressure. 
 	
 	It has been shown that annealing temperature correlates to mechanical loss in tantala thin films \cite{Vajente2018,Amato2018,Craig2015}. It may be that as the bubble diameters increase at higher annealing temperatures they begin to negatively affect the mechanical loss. It is also possible the bubbles may cause increased optical scattering and heat absorption in the films.
 	
 	The technique detailed in this paper could be applied to other gas bubbles in other materials. It may be of particular interest in the nuclear field where bubbles of noble gases often form in nuclear reactor walls as the mechanics are very similar to gas bubbles formed during annealing of films made using ion-beam deposition. 
	
	\section{Conclusion}
	
	Scanning transmission electron microscopy has been used to study the behaviour of argon in argon ion-beam assisted deposited thin films of $\text{Ta}_{2}\text{O}_{5}-\text{TiO}_{2}$ mixed oxides of the type used for multilayer mirrors in the Advanced LIGO gravitational wave detectors (and similar detectors elsewhere in the world such as advanced Virgo). It is shown using HAADF imaging that bubbles start to form after annealing at 400\;$^{\circ}$C and coarsen after annealing at 600\;$^{\circ}$C to a mean diameter of 22\;\AA. It was found that argon could be unambiguously detected in the films using EELS, and that this was concentrated in the bubbles in the sample annealed at 600\;$^{\circ}$C. Using an argon L$_{2,3}$ edge from an older study together with a Hartree-Slater calculation for the cross section well after the edge onset, a semi-empirical differential EELS cross-section was created and used for absolute quantification. This was then used to quantify the contents of 35 bubbles. It was found that these bubbles had a range of densities and pressures, with a mean density of 870\;kg/m$^3$ and a mean pressure of 400\;MPa. There was a weak trend towards lower pressure with increasing bubble diameter. It is currently unclear what influence the formation and coarsening of the bubbles has on the properties of the films, and whether changing the deposition process to remove the argon bubbles would be beneficial, although this should certainly be investigated in future work. Moreover, now it is clear that argon bubbles are present in most of these films, investigations including theoretical simulation-based studies should be conducted to understand their behaviour and whether they affect the optical or mechanical loss. Finally, the methodology used here for studying small noble gas bubbles is directly applicable to other noble gases in a range of materials, such as in cases of ion implantation.
	
	\section{Acknowledgements}
	
	RBC is supported by the Engineering  and  Physical  Sciences  Research  Council through grant EP/R513222/1. The authors would like to thank Martin Hart for making the samples and for Hafizah Noor Isa for her work on characterising the composition of the matrix of the samples. For helpful discussions, the authors would also like to thank Prof. Alan J. Craven and Prof. Stuart Reid. 
	
	This work was part of a programme of work funded in the UK by the STFC through the grants ST/I001085/1 and ST/N005422/1. This paper has LIGO DCC number P2000354.The data that support the findings of this study can be found in Enlighten: Research Data at 10.5525/gla.researchdata.1074.

\end{document}